\documentclass[conference]{IEEEtran}
\IEEEoverridecommandlockouts
\usepackage[utf8]{inputenc}
\usepackage{amsmath,amssymb,amsfonts}
\usepackage{algorithmic}
\usepackage{graphicx}
\usepackage{balance}
\usepackage[draft]{hyperref}
\usepackage{textcomp}
\usepackage{xcolor}
\usepackage{dblfloatfix}
\usepackage{enumitem}

\def\BibTeX{{\rm B\kern-.05em{\sc i\kern-.025em b}\kern-.08em
		T\kern-.1667em\lower.7ex\hbox{E}\kern-.125emX}}

\begin{document}
	
	\title{Scalable and Reliable Multi-Dimensional Aggregation of Sensor Data Streams
		\thanks{This research is funded by the German Federal Ministry of Education and Research (BMBF) as part of the Titan project (https://www.industrial-devops.org, grand no.\ 01IS17084B).}
	}
	
	\author{\IEEEauthorblockN{Sören Henning}
		\IEEEauthorblockA{\textit{Software Engineering Group} \\
			\textit{Kiel University}\\
			24098 Kiel, Germany \\
			 \enskip soeren.henning@email.uni-kiel.de \enskip}
		\and
		\IEEEauthorblockN{Wilhelm Hasselbring}
		\IEEEauthorblockA{\textit{Software Engineering Group} \\
			\textit{Kiel University}\\
			24098 Kiel, Germany \\
			\enskip hasselbring@email.uni-kiel.de \enskip}
	}
	\author{
	\IEEEauthorblockN{
		Sören Henning\IEEEauthorrefmark{1},
		Wilhelm Hasselbring\IEEEauthorrefmark{2}
	}
	\IEEEauthorblockA{
		Software Engineering Group, Kiel University, 24098 Kiel \\
		\IEEEauthorrefmark{1}soeren.henning@email.uni-kiel.de, 
		\IEEEauthorrefmark{2}hasselbring@email.uni-kiel.de
	}
	}
	
	\maketitle

	\begin{abstract}
	
	Ever-increasing amounts of data and requirements to process them in real time lead to more and more analytics platforms and software systems being designed according to the concept of stream processing. A common area of application is the processing of continuous data streams from sensors, for example, IoT devices or performance monitoring tools. In addition to analyzing pure sensor data, analyses of data for %
	groups of sensors often need to be performed as well. Therefore, data streams of the individual sensors have to be continuously aggregated to a data stream for a group.
	
	Motivated by a real-world application scenario, we propose that such a stream aggregation approach has to allow for aggregating sensors in hierarchical groups, support multiple such hierarchies in parallel, provide reconfiguration at runtime, and preserve the scalability and reliability qualities induced by applying stream processing techniques.
	We propose a stream processing architecture fulfilling these requirements, which can be integrated into existing big data architectures. We present a pilot implementation of such an extended architecture and show how it is used in industry. Furthermore, in experimental evaluations we show that our solution scales linearly with the amount of sensors and provides adequate reliability in the case of faults.
		
	\end{abstract}

	\section{Introduction}
	
	Stream processing \cite{Cugola2012, Roger2019} has evolved as a paradigm to process and analyze continuous streams of data, for example, coming from IoT sensors. The rapid development of stream processing engines \cite{Karimov2018} over the last years has paved the way for applications that process data exclusively online, i.e., as soon as it is recorded. Whereas a couple of years ago Lambda architectures were the de-facto standard for analytics platforms, currently more and more platforms follow the Kappa architecture pattern, where data is exclusively processed online \cite{Lin2017}. Further, entire software system architectures follow patterns such as asynchronously communicating microservices \cite{Hasselbring2017} and event sourcing \cite{Fowler2005}, which require data to be available as continuous streams instead of actively polled from databases.
	
	When considering continuous streams of measurement data, for example, from physical IoT sensors or software performance metrics, often an aggregation of multiple such streams is required. Whereas in the traditional approach of first writing all measurements to a (relational) database and then querying this database, this is a well-known task, performing such an aggregation on continuous data streams poses difficulties. This is in particular true when requirements for scalability and reliability have to be considered.
	
	In this paper, we contribute to the seminal work on stream processing by presenting and evaluating an approach to aggregate data of multiple streams in real time. The remaining paper is structured as follows: Section~\ref{sec:example} motivates the demand for our aggregation approach by an example for IoT sensor data streams. Section~\ref{sec:requirements} derives essential requirements for an aggregation approach. Section~\ref{sec:approach} presents our Titan stream processing architecture for aggregating data streams. Section~\ref{sec:case-study} shows how we implement this architecture in an industry-applied IoT monitoring platform. Section~\ref{sec:evaluation} evaluates our proposed architecture in terms of scalability and reliability.
	Section~\ref{sec:limitations} discusses limitations of our approach and possible extensions to overcome them.
	Finally, Section~\ref{sec:related-work} discusses related work and Section~\ref{sec:conclusions} concludes this paper.
	
	\section{Motivating Example}\label{sec:example}
	
	Operators of industrial production environments have high interest in getting detailed insights into the resource usage of machines and production processes. Those insights may reveal optimization potential, provide reporting for stakeholders, and can be used for predictive maintenance. Today's industrial production environments operate a variety of network-compatible measuring instruments. Such instruments (sensors) continuously measure, inter alia, the resource usage of individual machines, for example, their electrical power consumption. They may publish these metrics via a messaging system, allowing real-time analytics systems to collect, process, store, and visualize those data.
	
\begin{figure}[t]
		\centering%
		\includegraphics[width=\linewidth]{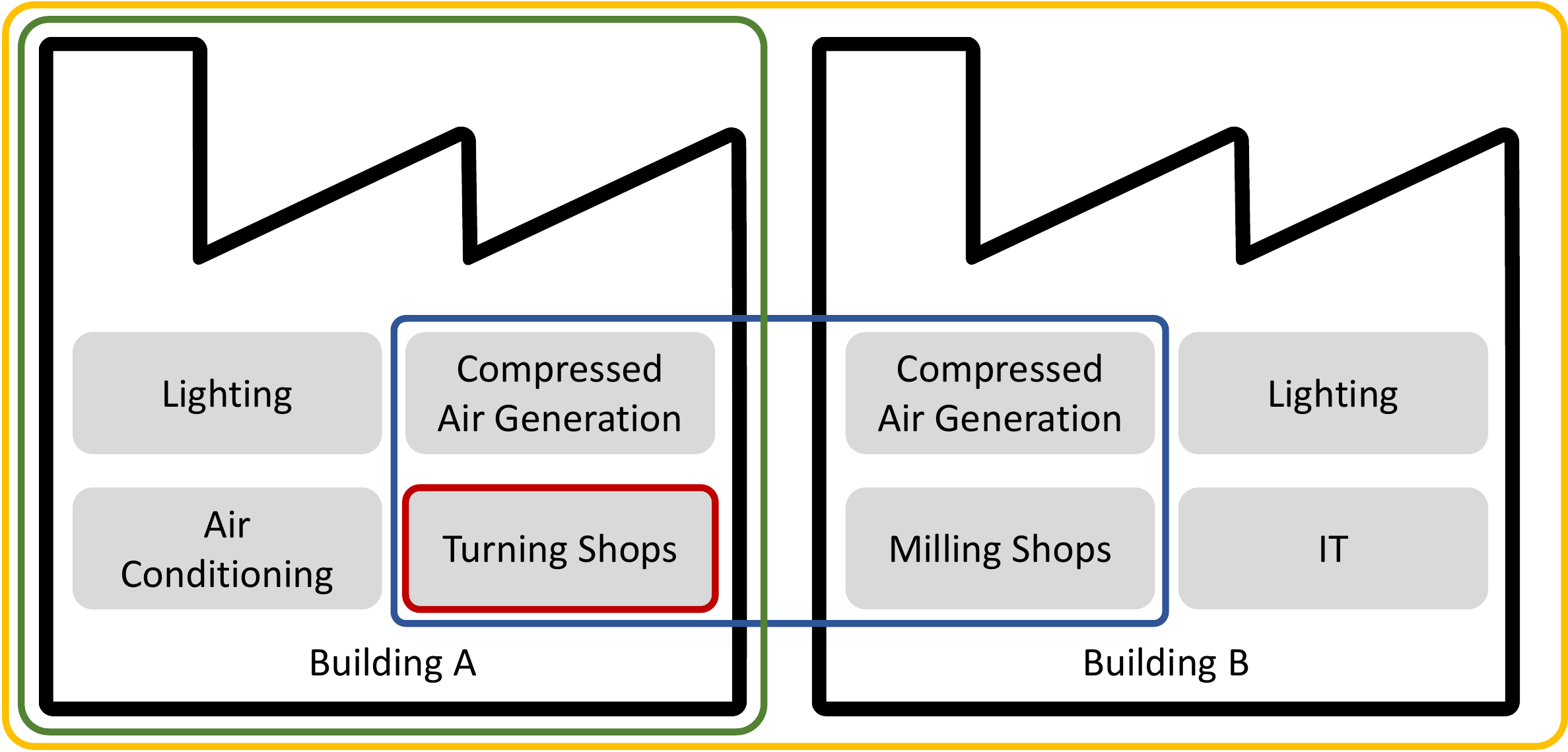}%
		\caption{Schematic illustration of a manufacturing company operating two buildings and a wide range of power-consuming machinery and infrastructure. Operators may be, for instance, interested in the total power consumption of all turning shops (red), directly required for production processes (blue), of a certain building (green), or of the entire company (yellow).}
		\label{fig:example}
	\end{figure}

	In particular, but not exclusively in the case of electrical power consumption, it is not only producing machinery that uses resources but also other company areas such as IT infrastructure, employee offices, or building technology. This leads to the situation that the amount of resource consuming devices is often immense, which makes it difficult to assess. Therefore, metrics for groups of machines in addition to consumption metrics of the individual machines are often required. Consequently, the data streams of the individual sensors have to be continuously aggregated.
	Referring to Fig.~\ref{fig:example}, which exemplary shows a production environment comprising various machines and devices, operators may require to answer questions such as:
	What is the resource usage of a certain machine type (e.g., the total power consumption of all turning shops)? What is the resource usage by business unit (e.g.,~the total power consumption of producing machinery)? What is resource usage by physically collocated machines (e.g., the total power consumption per building or shop floor)? What is the overall company-wide resource usage? Further, for machines featuring multiple independent power supplies (e.g., for redundancy), already obtaining consumption data for single machines requires to aggregate data streams of their individual power supplies.
	
	\section{Requirements for Stream Aggregation}\label{sec:requirements}
	
	Even though we motivated the need for real-time aggregation of data streams by resource data recorded by IoT devices, similar kinds of data aggregations are required by several other types of continuous sensor data streams. Continuing our motivating example, we derive the following requirements for a data stream aggregation architecture:
	
	\subsubsection{Multi-Layer Aggregation} Measurements of sensors are aggregated to groups of sensors. These groups can again be aggregated to larger groups and so forth. In the previous example, these could first be groups of individual machines of the same type, then groups of machines fulfilling the same function, then all machines used by the same production step and, further, all machines in the entire production environment.
	
	\subsubsection{Multi-Hierarchy Aggregation} In addition to a single hierarchy as described above, it is likely that there is a demand for supporting multiple such hierarchies.
	Referring to the previous example, besides a hierarchy based on the purpose of machines, one may also need a hierarchy which represents the physical location. For example, in a first step machines in the same shop floor are grouped and then all shop floors in a certain building are grouped.
	
	\subsubsection{Hierarchy Reconfiguration at Runtime} Stream processing applications are often characterized by demands for high availability.
	Therefore, we search for an approach that allows to modify or extend the previously described hierarchies at runtime to prevent reconfigurations causing downtimes.
	
	\subsubsection{Preserving Scalability and Reliability}
	Stream processing is usually used for large volume of data where the load has to be handled by multiple CPU cores or even multiple computing nodes.
	Therefore, an approach that performs aggregations on these streams has to preserve scalability and reliability properties to not become the overall architecture's bottleneck.

	\section{Our Titan Approach}\label{sec:approach}
	
	In this section, we present a generic approach to hierarchically aggregate streams of data. We start by a brief summary of the dual streaming model, which provides the foundation of our approach. Based on this model, we then introduce the actual approach in terms of a topology architecture.

	\subsection{The Dual Streaming Model}\label{sec:dual-streaming}
	
	The \textit{dual streaming model} \cite{Sax2018} is a model to define the semantics of a stream processing architecture. It adopts the notion of data streams and streaming operators from other established stream processing models \cite{Abadi2003, Akidau2013, Akidau2015}.
	
	A \textit{data stream} is an append-only sequence of immutable records, where records are key-value pairs augmented by a timestamp. Key-value pairs allow for data parallelization as records with different keys can be processed in parallel. This is the fundamental idea of building highly scalable stream processing applications.
	
	\textit{Streaming operators} are functions applied to each record of an input stream, whose results are appended to an output stream. The number of output records may be zero, one, or more than one depending on the type of operator. Usually, operators are distinguished between being stateless or stateful. Stateless operators produce an output solely based on the currently processed input record, whereas stateful operators may also take previous input records and computations into account.
	Successively connecting operator output streams with other operators allows to define complex stream processing topology architectures, for example, to build big data analytics applications.
	
	The dual streaming model extends these models by considering the result of streaming operators as successive updates to a table. These updates may be materialized into a versioned table or represented as a stream of insert, update, and delete events, inducing a duality of streams and tables.
	
	Sax et al. \cite{Sax2018} %
	present a reference implementation of the dual streaming model called Kafka Streams, a stream processing framework build on top of the distributed messaging system Apache Kafka \cite{Wang2015}. As in some cases the dual streaming model abstracts too many details to comprehensibly explain our topology, we use some architectural elements which are only present in their reference implementation, but not in the model.
	
	\subsection{Topology Architecture}
	
	\begin{figure*}[t]
		\centering%
		\includegraphics[width=\textwidth]{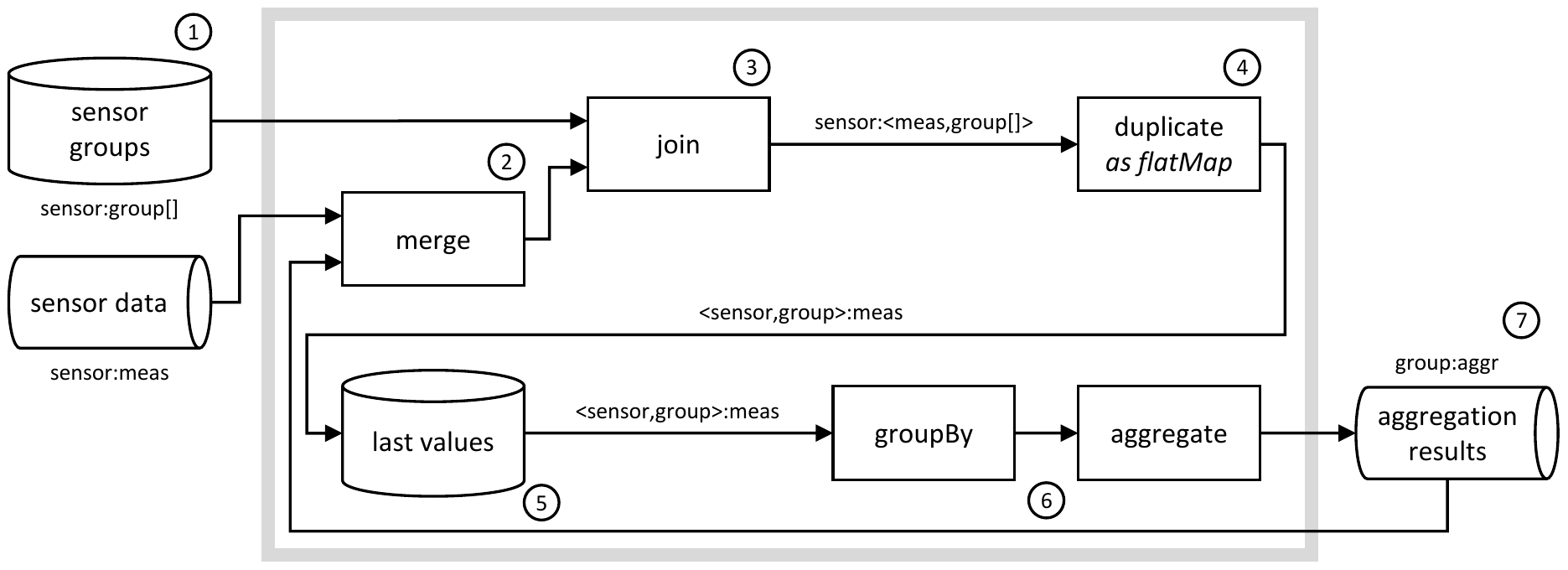}%
		\caption{Topology of our proposed stream processing architecture. Vertical cylinders represent tables whereas horizontal cyclinders represent data streams. Streaming operators are represented by rectangular boxes, which are connected to other operators, tables, or streams by directed arrows. Annotations at connections, tables, and streams indicate the corresponding type of contained or transmitted data, where key and value type are separated by a colon. \textit{sensor} represents a unique identifier for a sensor and \textit{group} such an identifier for a sensor group, \textit{group[]} represents a set of group identifiers, \textit{meas} represent a measurement, and \textit{aggr} represent an aggregation result. Everything located inside the gray box corresponds to the part of our approach, which can be deployed as an individual microservice. The  tables and streams placed outside can be considered as interfaces to other components.}
		\label{fig:architecture}
	\end{figure*}%
	
	We apply the dual streaming model to model the topology of consecutive operations on the streaming data, which are required for aggregating sensor data. This model forces an unidirectional and side-effect-free description of the data flow and, thus, allows the scalability and reliability facilitated by the model to be exploited. The architecture described in this way can be implemented as an encapsulated component, e.g., as a microservice, which can be integrated into existing software systems. Fig.~\ref{fig:architecture} visualizes our proposed topology architecture. The individual processing steps are described in the following.
	
	\subsubsection{Data Sources}
	Our proposed topology requires two data sources.
	The first one is an input stream of measurements, keyed by a sensor identifier. This is the sensor data stream as it comes, for example, from IoT devices or performance monitoring tools.
	The second one is a table, mapping sensors or sensor groups to the sensor groups containing them. A table entry consists of a key, which is the identifier of a sensor or group and a value, which is a set of all groups, this sensor (group) is part of. This table can, for example, be created from a stream, which captures changes in the hierarchy. It is not important which hierarchy a group belongs to, the only requirement is that identifiers are unique among multiple hierarchies.
	
	\subsubsection{Merging Measurement Streams}
	The first operator merges the input stream with a stream of already calculated aggregation results (see step 7).
	Note that the aggregated stream may require an additional converting step.
	In the following, we make no distinction between sensor measurements and aggregation results and call these values simply measurements.
	
	\subsubsection{Joining Measurement Stream and Group Table}
	The measurement stream and the groups table are joined using an inner join operation. This leads to a new update of the resulting table whenever either a measurement arrives or the groups a sensor belongs to changes. The result of this join operation is a tuple consisting of the measurement and the set of all groups this measurement has effect on.
	
	\subsubsection{Duplicating Join Results}
	In the next step, the measurements are duplicated in a way that for each group a new record is forwarded. This record has the following form: The key is a pair of sensor identifier and the corresponding group this record is created for. The value is the measurement. This operation is stateful as it always stores the last set of sensor groups and compares them with the new one. If a sensor was part of a group in a previous record but not in the currently processed one, a special record is forwarded with a ``tombstone'' value. This value serves for informing the following topology operators that the corresponding sensor is not longer part of the corresponding group.
	
	\subsubsection{Immediate Result: Last Value Table}
	The duplicated records are materialized to a table, which lists the last measured value per sensor and group. An arriving tombstone record for a sensor-group-pair deletes the corresponding entry in the table. This table of last values is the entry point for the following aggregation.
	
	\subsubsection{Grouping and Aggregating}
	Similar to an SQL group-by operation, table entries are grouped by their group name (second part of the key). The result is a grouped table containing one entry per group identifier. This table is then aggregated using appropriate adding and subtracting functions resulting in one aggregation result per group identifier.
	As defined by the dual streaming model, whenever an entry in the last values table is updated, a corresponding update record updates the grouped table and triggers the computation of a new aggregation result. This is done by first calling the subtract function for the previous record (e.g., subtracting the sensors previous measurement from the total group's value), followed by calling the add function for the new value (e.g., adding the sensors new measurement to the total group's value). Deleting an entry in the last value table (via a tombstone record) solely causes calling the subtract function.
	
	\begin{figure*}[t]
		\centering%
		\includegraphics[width=\textwidth]{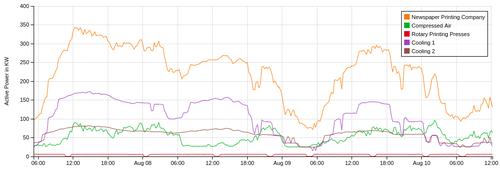}%
		\caption{Screenshot of the Titan Control Center's comparison view. It shows the overall power consumption of the \textit{Kieler Nachrichten Druckzentrum} in comparison to the main subconsumers. The overall consumption was continuously computed by aggregating the measurements of all subconsumers.}
		\label{fig:case-study}
	\end{figure*}%
	
	\subsubsection{Output: Aggregation Results}
	As a last step, the aggregated values per sensor group are published to a data stream. On the one hand, this stream serves as an interface so that other applications or services can use these data as they were real measurements. On the other hand, the stream is fed back to the beginning of the topology, where it can be used to compute aggregated values for sensor groups containing this group.

	\section{Industrial Case Study: IoT Sensor Data}\label{sec:case-study}
	
	In this section, we return to our motivating example from Section~\ref{sec:example} and show in a pilot implementation with our partner \textit{Kieler Nachrichten} how our proposed architecture can be used to aggregate IoT sensor data. The Titan Control Center \cite{Henning2019} is a microservice-based application for analyzing the energy consumption in manufacturing enterprises. It integrates energy consumption data of different data sources (e.g., machine-level data, building technology, or external software systems) and aggregates, analyzes, and visualizes them in near real time.
	
	We extended the Control Center's architecture by an additional microservice, which implements the topology described above and replaces a former, less scalable and reliable data aggregation. This microservice subscribes to a stream providing energy consumption data of individual machines and to a stream forwarding changes to the sensor hierarchy (published by the Configuration microservice of the Control Center).
	It aggregates the data of all sensors in a group by summing them up and publishes every result to a dedicated topic allowing other services to subscribe to this aggregated data. Other microservices use these data, for example, to calculate power consumption statistics, to produce forecasts, or to visualize it.
	
	Specifying the proposed architecture with the dual streaming model allows for a straightforward implementation in Kafka Streams. Nevertheless, as the dual streaming model is more abstract than the Kafka Streams API, we had to introduce some additions: In some places, streams had to be explicitly converted into tables via \textit{reduce} operations. The data type of aggregated records does not match the data type for sensor measurements, thus, an additional \textit{map} operation before merging with the sensor data stream was necessary. To avoid emitting subtract events from the table of aggregations, these have to be explicitly filtered out. The operator duplicating records for each parent had to be implemented as a custom \textit{flatTransform} operator, as Kafka Streams currently does not support \textit{flatMap} operations on tables.
	Using Kafka Streams promises a high degree of scalability and reliability, as deploying multiple instances of this microservice causes Kafka Streams to balance the load among all instance based on the topology and the number of configured Kafka topic partitions.
	
	Our implementation is used in production with two manufacturing enterprises to apply Industrial DevOps \cite{Hasselbring2019}. In these enterprises, the aggregation results are used, for example, to gain insights into how much energy is used by certain types of machines (e.g., the overall air conditioning), how big the difference is between measured company-wide energy consumption and the sum of all known consumers, or how much an individual machine contributes to the overall consumption of all machines of that type. Fig.~\ref{fig:case-study} shows a screenshot of the Titan Control Center's comparison view. It visualizes the total electrical power consumption of our partner \textit{Kieler Nachrichten Druckzentrum}, a newspaper printing company, in comparison to the consumption of its major consumers.

	\section{Experimental Evaluation}\label{sec:evaluation}
	
	We experimentally evaluate the scalability and the reliability of the proposed stream processing architecture.
	For these evaluations, we simulate different numbers of power metering sensors and aggregate their data streams to pre-defined groups using the Titan Control Center (see above). Each simulated sensor emits one measurement per second. We record both the number of sensor measurements per sensor and the number of computed aggregation results per second as well as the average latency per second between sensor record generation and obtaining the result of an aggregation.
	The experimental setup is deployed in a Kubernetes cluster of 4 nodes each equipped with 384~GB RAM and $2\times16$ CPU cores providing 64~threads (overall 256~parallel threads).

	\subsubsection{Evaluation of Scalability}
	
	In order to assess the scalability of our proposed approach, we evaluate how an increasing amount of input data can be handled by an increasing number of aggregation instances. For this purpose, we determine the number of instances required to aggregate the data streams of a given workload. We group 8 simulated sensors into one group and group 8 of such groups again into one larger group. We evaluate 4 workloads, where we simulate 2, 3, 4, or 5 nested groups, resulting in a total amount of sensors of $8^2$, $8^3$, $8^4$, and $8^5$ and a corresponding number of records per second. For each scenario, we deploy different numbers of instances ranging from 1 to 128.
	
	We consider a deployment (i.e., a certain set of deployed instances) as being sufficient to handle the given workload if all simulated data can be processed such that no data records are piling up between their generation and the aggregation. To obtain this information, we measure the latency between generation and aggregation. If this latency remains reasonably constant, we conclude that records are aggregated at approximately the same speed as they are generated. We apply linear regression to compute a trend line and consider the processing to be reasonably constant if the trend line's slope is less than 100~ms.\footnote{Due to Kafka Streams' task model, the throughput is subject to large fluctuations. The calculated trend line can therefore be inaccurate, suggesting a more conservative threshold for the slope of 100~ms. Lower threshold return similar median results, but produces more outliers.}
	For each evaluated workload, we determine the minimum number of instances, which is required to handle that workload, i.e., shows a latency trend line with a slope of less than 100~ms. This evaluation is repeated 10 times.
	
\begin{figure}[bt]
		\centering%
		\includegraphics[width=\linewidth]{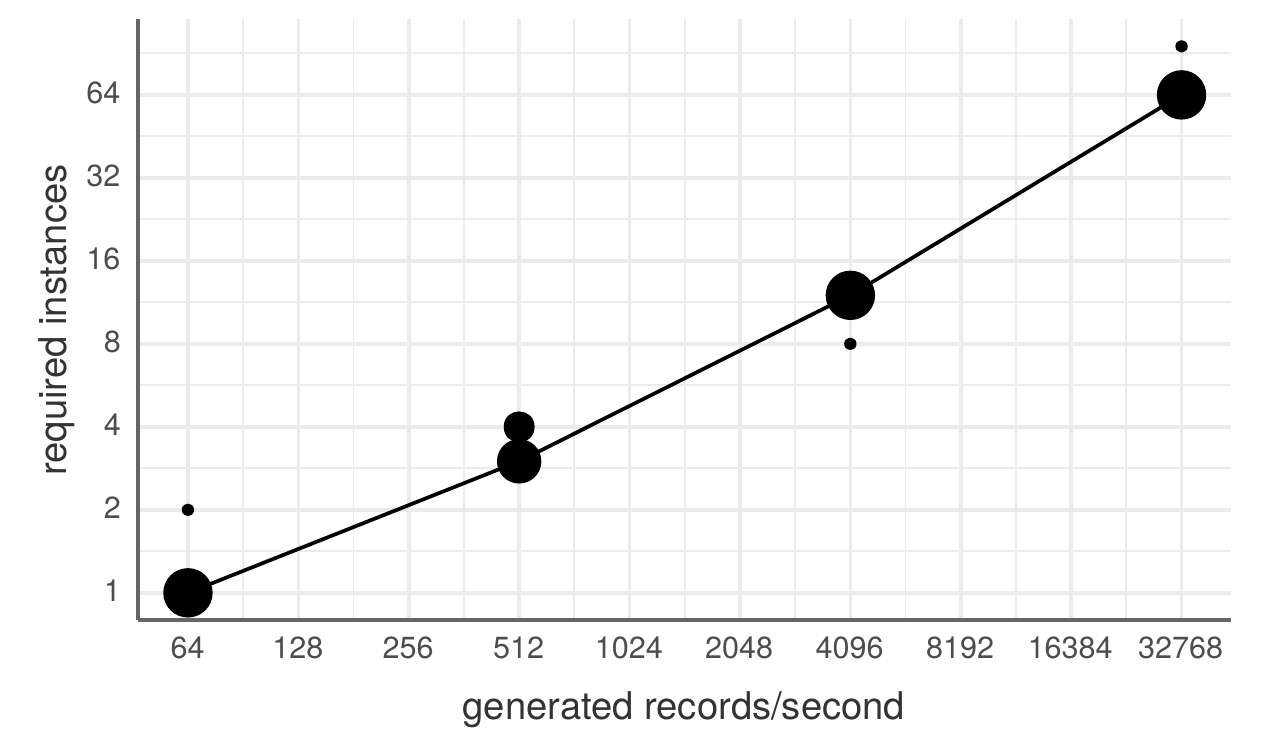}%
		\caption{Number of required instances for aggregation in relation to the number of generated records per second. The size of points indicates the frequency of the respective observation. The black line connects the median numbers of required instances per workload.}
		\label{fig:scalability}%
	\end{figure}

	Fig.~\ref{fig:scalability} shows the median over all repetitions of the required number of instances per workload. We notice that the required number of instances scales linearly with the amount of data to be aggregated.
	
	\subsubsection{Evaluation of Reliability}

	In order to assess the reliability of our proposed approach, we evaluate how the architecture behaves when components fail. We generate the workload described in the scalability evaluation, which simulates 4 nested groups of sensors, and aggregate the data with 24 instances of the aggregation component. After 10~minutes of processing, we inject a failure during operation by shutting down 18~instances and starting them again after 5~more minutes. We measure both the number of generated messages and the amount of aggregation results during the entire evaluation. Since both values fluctuate strongly, we additionally calculate a moving average over a 60~seconds window.
	
\begin{figure}[bt]
		\centering%
		\includegraphics[width=\linewidth]{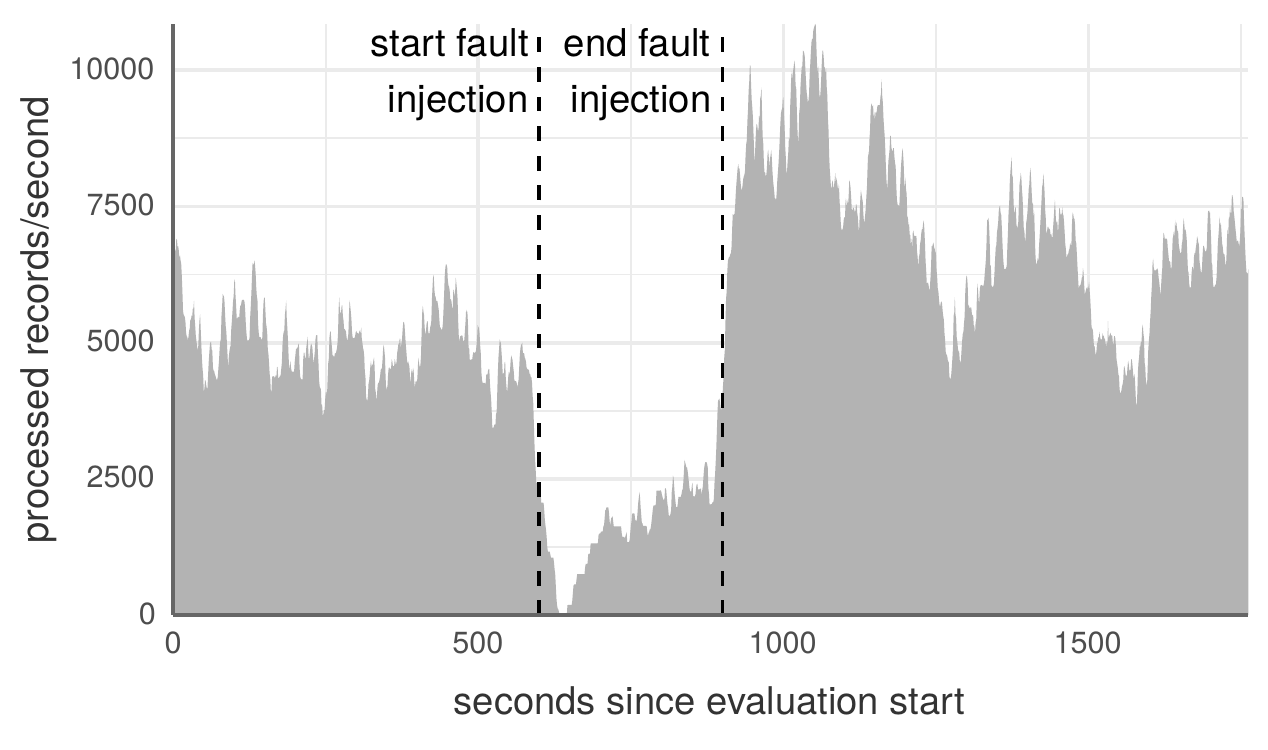}%
		\caption{Number of processed records per second over a moving average of 60~seconds in the course of time. The two dashed vertical lines indicate the point in time at which the simulated failure starts or ends.}
		\label{fig:reliability}%
	\end{figure}

	The average number of processed records per second over time is presented in Fig.~\ref{fig:reliability}. It can be seen that the amount of performed aggregations decreases sharply during the simulated failure. This is reasonable as there are not enough resources available anymore to process all data. However, if the stopped instances are replaced by new instances, the amount of processed data increases again. Furthermore, as we set the number of processing instances twice the number actually necessary (cf. scalability evaluation), the piled up data is also processed.
	
	\section{Limitations and Possible Extensions}\label{sec:limitations}
	Our proposed architecture has two limitations: First, it yields incomplete results for the first records until every sensor generated at least one measurement. This could be fixed by adding additional logic in the flat transform step or for many use cases simply be ignored as after a short warm-up the approach would work as expected.
	
	The second limitation concerns out-of-order records. According to the dual streaming model, a late arriving record would cause a recomputation of all consecutive steps. While this is in principle desired, the corresponding lookup in the last values table only yields the most recent value causing wrong results. A simple, yet in many cases sufficient solution is to simply reject late arriving records. A more complex, but also correct solution is to introduce windowing operations, which compute aggregation results for time windows of, e.g., one minute.
	In effect, this approach would maintain not only one last values table but instead one last values table per time window.

	\section{Related Work}\label{sec:related-work}

	Joins in stream processing (analogous to those in SQL) \cite{Arasu2006}	allow to connect different or identical streams. As the join operation is a bivariate %
	function and thus aggregating multiple streams would require a chain of join operations, a respective pipeline can only statically be created and not dynamically adjusted at runtime.
	This contrasts our approach, which allows reconfigurations at runtime by changing the sensor groups table or an underlying stream.

	Another type of aggregating data in streams is along the dimension of time. In this case, records within the same time window having the same key are aggregated to a new record. Depending on the actual requirements, records can be aggregated within fixed windows, sliding windows, or session-based windows \cite{Akidau2015}. In addition to research on the efficient computation of window aggregations \cite{Li2005, Traub2019}, there are also publications proposing software architectures for a temporal aggregation platform, for example by Twitter \cite{Yang2018}.
	Temporal aggregations are compatible with our approach, which means that temporal aggregated streams of sensor data can be further aggregated to groups and streams for sensor groups can be further aggregated over time.
	
	As most approaches applying stream processing, our presented approach is  primarily based on data parallelism, meaning that (sub)topologies exist in multiple instances, each processing a portion of the data.
	Pipe-and-Filter frameworks such as TeeTime \cite{Wulf2017} employ task parallelism, where the individual filters (operators) are executed in parallel.
	Since in this way all data pass each filter, the identified requirements can be realized in a single filter. In contrast to the solution we presented, however, scalability would be significantly compromised.
	
	\section{Conclusions}\label{sec:conclusions}
	
	Software systems, which analyze or react on sensor data streams, often have to process not only raw measurements but also aggregated data for groups of sensors. In this paper, we presented our Titan approach for continuously aggregating sensor data. It supports the aggregation of hierarchical groups, multiple such groups in parallel, and %
	reconfigurations at runtime. For this purpose, we designed a stream processing architecture, which can be integrated (e.g., as a microservice) into existing big data architectures. It consists of a topology of stream processing operators, requires two input data streams, which provide sensor group hierarchies as well as the sensor data stream, and provides an output stream of aggregated data. We provide an implementation of this architecture for power consumption data and show how our implementation can be integrated into an analytics platform used in industry. Furthermore, in an experimental evaluation, we show that our proposed architecture scales linearly with the amount of sensors and tolerates faults during operation.
	A replication package and experimental results provided as supplemental material allow to repeat and extend our work \cite{Henning2019b}.
	
	For future work, we plan to add optional support for out-of-order records by aggregating measurements in time windows as described in Section~\ref{sec:limitations}.
	Furthermore, future work may explore how our architecture can be implemented with other stream processing engines, for example, by considering recent trends towards uniform stream query languages \cite{Begoli2019}.
	As we were not able to discover any scalability limitations in our conducted experimental evaluation, we also plan to conduct experiments with even larger amounts of sensors.

	\bibliographystyle{IEEEtran}
	\bibliography{IEEEabrv,references}

\end{document}